\documentclass[%
 reprint,
superscriptaddress,
 amsmath,amssymb,
 aps,
prb,
floatfix,https://www.overleaf.com/project/60e89d41b906ae79602fa7f4
]{revtex4-2}
\usepackage{graphicx}
\usepackage{dcolumn}
\usepackage{bm}
\usepackage{subfig}
\usepackage[font=small]{caption}
\usepackage{svg}

\begin{document}

\preprint{APS/123-QED}

\title{Microscopic design of a topologically protected singlet-triplet qubit in an InAsP quantum dot array}

\author{Jacob Manalo}
\affiliation{%
 University of Ottawa, Ottawa, Canada K1N 6N5 
}

\author{Daniel Miravet}%
\affiliation{%
 University of Ottawa, Ottawa, Canada K1N 6N5 
}

\author{Pawel Hawrylak}
\affiliation{%
 University of Ottawa, Ottawa, Canada K1N 6N5 
}%

\date{\today}

\begin{abstract}
We present here the steps enabling the microscopic design of a topologically protected singlet-triplet qubit in an InAsP quantum dot array embedded in an InP nanowire. The qubit is constructed with two Haldane spin-$\frac{1}{2}$ quasiparticles in a synthetic spin one chain. The qubit is described by a two-leg multi-orbital Hubbard Kanamori (HK) model with parameters obtained from the microscopic calculations of up to eight electrons in a single and double quantum dot. In this HK model describing long arrays of quantum dots, using both exact diagonalization and matrix product state (MPS) tools, we demonstrate a four-fold quasidegenerate ground state separated from excited states by a finite energy gap similar to a Heisenberg spin-1 chain in the Haldane phase. We demonstrate the existence of spin-$\frac{1}{2}$ quasiparticles at the edges of the chain by observing the magnetic field dependence of the low energy spectrum as a function of applied magnetic field. The applied magnetic field also isolates the singlet and $S^z=0$ triplet states from the other triplet components allowing these states to serve as a qubit basis. Most importantly, the regions in parameter space where the low energy spectrum of the multi-orbital Hubbard chain yields a Heisenberg spin-1 chain spectrum are mapped out. Due to the finite energy gap, this qubit has the potential to be protected against perturbations.
\end{abstract}

\maketitle


\section{Introduction}

The development of solid state quantum information processing devices is currently a research area of great interest \cite{korkusinski_worldscientific_2008,deep_am_2022,loss_arcmp_2013, northup_nature_2014, lafferiere_apl_2021}. At the moment,  qubits developed for commercial use are superconducting \cite{johnson_nature_2011, gambetta_nature_2017,arute_nature_2019}, trapped ion \cite{monroe_nature_2019, monz_prx_2021},  electron spin \cite{brum_hawrylak_sm1997,loss_divincenzo_pra1998,simmons_nature_2021,vandersypen_nature_2006} and photonic qubits \cite{knill_nature_2001,madsen_nature_2022,milburn_rmp_2007} due to their robustness and scalability \cite{myerson_prl_2008, hassler_njp_2011, knill_nature_2001,mikel_prr_2021,madsen_nature_2022}. However, these qubits suffer from decoherence and a quest for topologically protected qubits continues \cite{jaworowski_appliedsci_2019,dassarma_nature_2012,marcus_nature_2018}. Recently, a qubit  constructed with two Haldane spin-$\frac{1}{2}$ quasiparticles \cite{haldane_physlett_1983, haldane_rev_mod_phys_2017, affleck_prl_1987} in a synthetic spin one 
chain  \cite{jaworowski_scirep_2017,jaworowski_appliedsci_2019} has been proposed \cite{shim-hawrylak-ssc2010,jaworowski_scirep_2017, hsieh_iop_2012}. A synthetic spin one chain could be realised using gated triple quantum dots \cite{shim-hawrylak-ssc2010}  , array of semiconductor quantum dots in a nanowire \cite{jaworowski_scirep_2017, manalo_prb_2021} and  a chain of triangular graphene quantum dots \cite{devrim_springer_2014, fasel_nature_2021, rossier_prb_2022}. Here, we discuss the atomistic design of a synthetic spin-1 chain using a semiconductor quantum dot array in a nanowire to realize a topologically protected singlet-triplet qubit.

Previous effective mass and Heisenberg model based spin calculations suggested that such a qubit can be realized using a chain of InAsP semiconductor quantum dots with 4 electrons each in a InP nanowire \cite{jaworowski_scirep_2017, jaworowski_appliedsci_2019,manalo_prb_2021}. Furthermore, it has been shown through microscopic calculations that the ground state of a single InAsP quantum dot in a nanowire is a spin triplet and that the low energy spectrum of an array of two InAsP quantum dots in an InP nanowire is similar to the spectrum of a Heisenberg chain of two spin-1 particles \cite{manalo_prb_2021}. The parameters of this two site Heisenberg Hamiltonian were used to extend the Heisenber spin-1 chain. 

Here, instead of effective Heisenberg Hamiltonian we derive and use an effective multi-orbital Hubbard model with parameters obtained from microscopic atomistic calculations. We determine a set of microscopic parameters for which a long macroscopic quantum dot chain with 4 electrons each has a four-fold quasi-degenerate ground state separated from the quintuplet state by a finite energy gap, similar to a Heisenberg spin-1 chain. We also show that the electrons in a quantum dot array behave the same way as two coupled spin-$\frac{1}{2}$ quasi-particles would in a magnetic field. 

Furthermore, we show that the length of array controls the singlet-triplet splitting while  the Zeeman splitting of the nonzero spin triplet states allows us to isolate the quasi-degenerate singlet and triplet states from the quintuplet allowing the isolated states to serve as a qubit basis. We then demonstrate that the multi-orbital Hubbard parameters which result in a Heisenberg spin-1 chain model form distinct regions in parameter space and not sporadic regions. Determining these parameters allows for the fine tuning of the spectral gap. The parameters are modified by controlling the size and As concentration of the InAsP quantum dot as well as interdot distance and material of the quantum dot array enabling  construction
of robust topologically protected singlet-triplet qubits.

The paper is organized as follows. First, we define the multi-orbital Hubbard model in terms of individual quantum dots and the interaction between them. We then describe the methodology of the calculations which include exact diagonalization and density matrix renormalization group (DMRG) in the formalism of matrix product states (MPS) \cite{white_prl_1992,schollwock_rev_mod_phys_2005,schollwock_rev_mod_phys_2005, verstraete_prl_2004}. Next, we analyze the low energy spectrum as a function of array size and show the behaviour of the chain in a magnetic field. Finally, we map out regions in parameter space where the multi-orbital Hubbard model gives a low energy spectrum that resembles that of the Heisenberg spin-1 chain.

\section{I\lowercase{n}A\lowercase{s}P Quantum Dot Array in an I\lowercase{n}P nanowire}
\begin{figure}
    \centering
    \includegraphics[width=\linewidth]{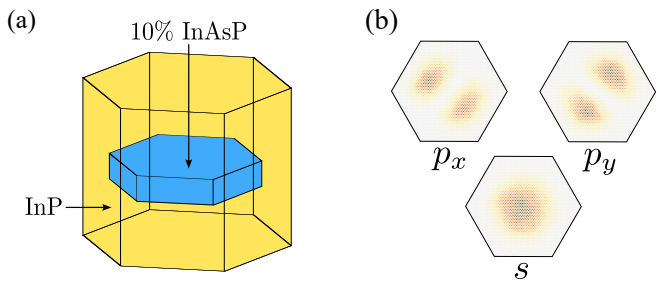}
    \caption{(a) Hexagonal InAsP quantum dot (blue) in an InP nanowire (yellow). (b) Charge densities of single particle states.}
    \label{fig:schematic}
\end{figure}

We aim to realize a synthetic spin-1 chain with an array of InAsP quantum dots embedded in an InP nanowire. The quantum dot array is constructed with a single InAsP quantum dot shown in Fig. \ref{fig:schematic}(a) as a building block. It has been shown that a synthetic spin-1 object is formed when 4 electrons are injected into the InAsP quantum dot \cite{manalo_prb_2021}. With each InAsP quantum dot acting as a spin-1 object, we construct a synthetic spin-1 chain with an array of these InAsP quantum dots as shown in Fig. \ref{fig:chain_diagram}.

The microscopic calculations for one and two InAsP quantum dots embedded in an InP nanowire serve as the foundation for the effective multi-orbital Hubbard model that describes the InAsP quantum dot array. Essentially, the microscopic calculations begin with a tight-binding model \cite{zielinski_prb_2010,sheng_prb_2005,cygorek_prb_2020,manalo_prb_2021} where the quantum dot nanowire is created by first building an InP matrix and defining a hexagonal nanowire inside as shown in Fig. \ref{fig:schematic}(a) where random P atoms are replaced with As atoms at a concentration of $10$\%. Fig. \ref{fig:schematic}(b) shows the probability densities of the single particle states obtained from the tight-binding model. Despite the random distribution of As atoms, the spectrum consists of an $s$-shell followed by two states of a $p$-shell.


Furthermore, it was shown that when four electrons were inserted into the quantum dot, two of the electrons filled the $s$-shell, leaving the other two electrons to form a triplet state on the $p$-shell. The many-body calculations of the $N$ electron complex were done using the configuration interaction method for the Hamiltonian given by
\begin{align}
    H_{MB} = \sum_i E_ic_i^\dagger c_i + \frac{1}{2}\sum_{ijkl}\langle ij|V| kl \rangle c_i^\dagger c_j^\dagger c_k c_l
    \label{eq:mb_ham}
\end{align}
where $E_i$ is the energy of single particle state $i$, $c_i^{\dag}$ ($c_i$) is the creation (annihilation) operator for an electron on state $i$ and $\langle ij|V| kl \rangle$ is the Coulomb matrix element where two electrons, one in state $i$ and another in state $j$ scatter to states $k$ and $l$. Likewise, the many-body spectrum of two quantum dots, each with four electrons, resembled the spectrum of two coupled spin-1 particles. The limitation of these microscopic calculations extended to a long chain of quantum dots is that they are computationally expensive. Computing such arrays where each quantum dot contains millions of atoms, with each atom containing $20$ spin-up and spin-down  orbitals, and 4 electrons per quantum dot is not possible. Since the singlet-triplet qubit requires a chain of many synthetic spin-1 quasiparticles, it is necessary to use a  simplified model that still captures the physics of a spin-1 chain.


\section{The Multi-Orbital Hubbard Model}

\begin{figure}
    \centering
    \includegraphics[width=\linewidth]{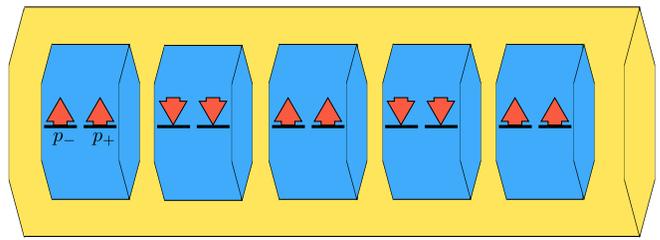}
    \caption{A chain of InAsP quantum dots (blue) embedded in an InP nanowire (yellow). 
    Red arrows indicate electrons with corresponding spin.}
    \label{fig:chain_diagram}
\end{figure}

We now turn to the effective multi-orbital Hubbard model to describe the quantum dot array shown in Fig. \ref{fig:chain_diagram}. In this model, each quantum dot is described as a site with two $p$ orbitals, $p_-$ and $p_+$. Here, $s$-shell electrons are ignored because the probability of $s$ electrons scattering to the $p$-shell is negligible due to the large $s$-$p$ splitting in the microscopic single-particle spectrum. Exchange interaction of additional two electrons  half-filling the $p$-shell can ferromagnetically couple their spins to form a synthetic spin-1 state as shown schematically in Fig. \ref{fig:chain_diagram}. 

To retain essential microscopic description of the quantum dot  we reduce the microscopic Hamiltonian in (\ref{eq:mb_ham}) to the effective multi-orbital Hubbard Hamiltonian for a single quantum dot given below 

\begin{align}
    H_0(i) & = U_1 \sum_\alpha n_{i\alpha\uparrow}n_{i\alpha\downarrow}+(U_2-\frac{J_{1/2}}{4})n_{i-}n_{i+} \nonumber \\
   & -J_{1/2} \mathbf{S}_{i-}\cdot\mathbf{S}_{i+} + \frac{\Delta}{2}\sum_\sigma \sum_{\alpha \neq \beta}c^{\dag}_{i\alpha \sigma} c_{i\beta \sigma}
    \label{eq:hk_singledot}
\end{align}
where $\alpha, \beta \in \{-, +\}$ denote the orbital indices, $\sigma \in \{\uparrow, \downarrow\}$ denotes spin and $n_{i\alpha}\equiv \sum_{\sigma} n_{i \alpha \sigma}$ is the number of electrons in orbital $\alpha$ in quantum dot $i$. This Hamiltonian as well as the Hamiltonian for a chain of quantum dots is derived from the microscopic Hamiltonian by employing certain approximations to the Coulomb matrix elements as described in Ref. \cite{manalo_prb_2021}. 

The first term is the Hubbard term, which describes the energy $U_1$ required for a spin-up and spin-down electron to occupy a single orbital. The second term describes the coupling between electrons on the $p_-$ and $p_+$ shells with energy $(U_2 - J_{1/2}/4)$. Here, $U_2$ is the direct Coulomb interaction between an electron on $p_-$ and an electron on $p_+$ and $J_{1/2}$ is the exchange between them. In general, $U_1$ and $U_2$ differ in value, but for the systems we are interested in, $U_1=U_2\equiv U$. The following $J_{1/2}$ term describes the Heisenberg ferromagnetic coupling between $p_-$ and $p_+$ electrons, which is not to be confused with the effective Heisenberg spin-1 coupling between quantum dots, hence the subscript $1/2$ in the coupling constant $J_{1/2}$. This spin-$\frac{1}{2}$ coupling arises from the exchange interaction between electrons on different orbitals. Finally, the last term describes the $p$-shell splitting due to the broken lateral symmetry of the quantum dot from the random distribution of As atoms, where the energy splitting $\Delta$ is the splitting between the $p_x$ and $p_y$ orbitals which are both linear combinations of $p_-$ and $p_+$.

To compute the spectrum of a chain of quantum dots, we must include the interaction between the quantum dots. The total multi-orbital Hubbard Hamiltonian is given by
\begin{align}
    H & = \sum_i \left( H_0(i) + t \sum_{\alpha \sigma}(c^{\dag}_{i\alpha\sigma}c_{i+1\alpha\sigma} +\textrm{h.c.}) \right) \nonumber \\&+ V\sum_i n_i n_{i+1} 
    \label{eq:hk_multidot}
\end{align}
which is the sum of all single quantum dot Hamiltonians in the array and the interactions between nearest neighboring dots. The first term that describes the interdot interactions is the tunneling term $t c^{\dag}_{i\alpha\sigma}c_{i+1\alpha\sigma}$ which describes the process of an electron hopping from quantum dot $i$ to the nearest neighbor quantum dot $i+1$ with a hopping energy $t$. The second term of the interdot interaction portion of the Hamiltonian describes the electrostatic interaction between electrons on neighboring quantum dots. Here, $n_i \in \left[ 0, 4 \right]$ is the electron occupation of dot $i$ and $V$ is the Coulomb matrix element which is direct with respect to dot index and is defined to be $V\equiv \langle i\alpha,j\beta | V| j\beta, i\alpha \rangle$ where $\alpha,\beta \in \{ p_+,p_- \}$.

The two most important terms in determining the behaviour of the system as a spin-1 chain are the intradot exchange term $-J_{1/2} \sum_i \mathbf{S}_{i-}\cdot\mathbf{S}_{i+}$, which describes the spin-spin coupling between a $p_-$ electron and a $p_+$ electron, and the tunneling term $ t\sum_i \sum_{\alpha \sigma}c^{\dag}_{i\alpha\sigma}c_{i+1\alpha\sigma}$. The intradot exchange term, which controls the electronic behaviour of the quantum dot as a spin-1 object is compromised by the tunneling term, which breaks the spin-$1$ apart. Without the interdot tunneling term however, the singlet, triplet and quintuplet states of the quantum dot array will all be degenerate which means that there is no finite gap in the spectrum.

The single dot parameters $U_1$, $U_2$, $J_{1/2}$, $\Delta$ and multidot parameters $t$ and $V$ were obtained by fitting the spectrum of (\ref{eq:hk_multidot}) to the microscopic tight-binding spectrum of a two dot array using a genetic algorithm. This effective multi-orbital Hubbard model with parameters obtained from microscopic calculations allows us to simulate an array with many dots so that we can construct the topologically protected singlet-triplet qubit.

\section{Methodology}

In this work, we compute the many-body spectrum of a large chain of quantum dots using the multi-orbital Hubbard model and Configuration Interaction and DMRG tools to demonstrate the similarity to the spectrum of a spin-1 chain with two Haldane spin-$\frac{1}{2}$ quasiparticles. Next, we apply a magnetic field to the quantum dot chain to determine the behaviour of the Haldane spin-$\frac{1}{2}$ quasiparticles at the edges as well as to find the magnetic field at which the qubit can operate. Finally, regions in parameter space, i.e. the parameters in (\ref{eq:hk_singledot}) and (\ref{eq:hk_multidot}), where the Hubbard chain produces a Heisenberg spin-1 chain spectrum are mapped out.

All calculations of spectra of arrays with two quantum dots are done with exact diagonalization while calculations of spectra of larger arrays are done with the DMRG algorithm \cite{white_prl_1992,schollwock_rev_mod_phys_2005,schollwock_rev_mod_phys_2005, verstraete_prl_2004}. In this work, we used iTensor and a tool that we developed called Python MPS (PyMPS) to perform the DMRG calculations \cite{itensor,pymps}. 

\begin{table}
\caption{\label{tab:params} Multi-orbital Hubbard parameters for the quantum dot chain}
\begin{ruledtabular}
    \begin{tabular}{cc}
         Parameter & Value (meV) \\
         \colrule
          $U$ & $15.971$ \\
         $J_{1/2}$  & $5.000$ \\
         $\Delta$  & $0.844$ \\
         $t$ & 2.389 \\
         $V$ & 8.05 \\
    \end{tabular}
\end{ruledtabular}
\end{table}

\begin{figure}
    \centering
    \includegraphics[width=0.9\linewidth]{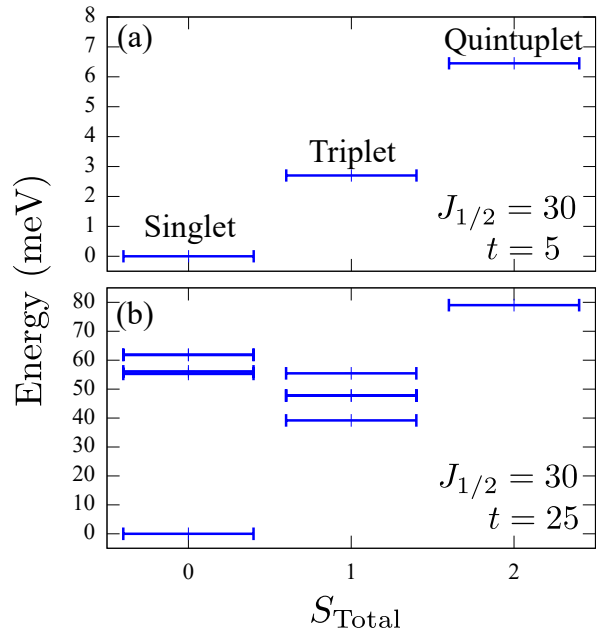}
    \caption{Low energy spectra of two quantum dots with two different sets of parameters. The spectrum (a) shows the spin-1 spectrum criterion satisfied, while (b) is an example where the criterion is not satisfied. All parameters are in meV. }
    \label{fig:2dot_spec_comparison}
\end{figure}

Table \ref{tab:params} shows the multi-orbital Hubbard model parameters that were obtained from microscopic calculations \cite{manalo_prb_2021}. An important feature of the multi-orbital Hubbard chain is the similarity of its low energy spectrum and the spectrum of a spin-1 chain. However, the similarity  to the spin-1 chain spectrum is dependent on the choice of multi-orbital Hubbard parameters. This is evident with the example of a chain of two quantum dots. The spectrum of the two quantum dot array is shown in Fig. \ref{fig:2dot_spec_comparison}(a) where the parameters except for $J_{1/2}$ and $t$ are taken from Table \ref{tab:params}. While the Heisenberg spin-1 chain spectrum is reproduced with the parameters shown in Fig. \ref{fig:2dot_spec_comparison}(a), it is not reproduced when those parameters are changed as shown in Fig. \ref{fig:2dot_spec_comparison}(b). The dependence of the spectrum on parameters allows us to define a criterion for Heisenberg spin-1 chain behaviour. The criterion is such that when the multi-orbital Hubbard spectrum replicates the spin-1 chain spectrum as shown in Fig. \ref{fig:2dot_spec_comparison}(a), the criterion is satisfied, otherwise, as illustrated in Fig. \ref{fig:2dot_spec_comparison}(b) when the spectrum of the spin-1 chain is not replicated, the criterion is not satisfied. This principle applies to long arrays of quantum dots. For $L=50$ quantum dots the $S^z=0$ Hilbert space of the HK model at half-filling is $\binom{100}{50}^2 \approx 10^{58}$. For such a large Hilbert space we apply MPS-DMRG tools to obtain the low energy spectrum. Fig. \ref{fig:50_dots}(a) shows an example where a chain of 50 quantum dots satisfies the spin-1 chain criterion. This criterion applies to any size of quantum dot array and will be imperative to mapping out regions in parameter space where the system behaves as a chain of spin-1s.

The low energy spectrum of the long chain shown in Fig. \ref{fig:50_dots}(a) illustrates the behaviour of two uncoupled spin-$\frac{1}{2}$ quasiparticles. While Fig. \ref{fig:2dot_spec_comparison}(a) shows that the two dot array resembles two spin-1s, Fig. \ref{fig:50_dots}(a) shows that the chain of many quantum dots resembles a chain of many spin-1 particles, which is understood in terms of two Haldane spin-$\frac{1}{2}$ quasiparticles at the edges. To illustrate the spectral gap, the spectrum of the multi-orbital Hubbard Hamiltonian as a function of system size shown in Fig. \ref{fig:50_dots}(b) was computed and compared with that of the Heisenberg spin-1 chain.

The spectrum of the Heisenberg spin-1 chain was computed using the Heisenberg Hamiltonian given by

\begin{align}
    H = J_{1}\sum_i \mathbf{S}_i \cdot \mathbf{S}_{i+1}
\end{align}
where $J_1=2t^2/\left( U+\frac{J_{1/2}}{2}-V \right)$ is the effective Heisenberg spin-1 coupling which is analytically obtained by treating the tunneling term in (\ref{eq:hk_multidot}) as a perturbation \cite{blazej_thesis_2018}.

We then added the term $g\mu B S_{\textrm{Total}}^z$ to (\ref{eq:hk_multidot}) to study the behaviour of the quantum dot array as a function of applied magnetic field. Using a chain of $20$ and $50$ quantum dots, we determined the array size required for the singlet-triplet splitting in Figs. \ref{fig:b_field}(a) and (b) to be small enough to avoid unwanted level crossings. 

Finally, to determine the set of multi-orbital Hubbard parameters where the array gives a Heisenberg spin-1 chain like spectrum, we set the following criterion; if the spectrum consists of a singlet ground state, followed by a triplet first excited state, then followed by a quintuplet second excited state with no other states in between, then the criterion is satisfied.

Fig. \ref{fig:2dot_spec_comparison} shows an example of a spectrum that satisfies the Heisenberg spin-1 criterion and another example that does not. In Fig. \ref{fig:2dot_spec_comparison} (a), $t = \frac{1}{6}U$, which is still in the perturbative regime, while in Fig. \ref{fig:2dot_spec_comparison} (b), $t \sim U$ thus the spin-1 description is no longer valid. Unlike the Heisenberg spin-1 chain spectrum, there are intermediate singlet and triplet states that appear below the quintuplet energy due to the coupling of the ground state singlet and triplet to the higher energy configurations that contain triple electron occupation in a dot \cite{manalo_prb_2021}. We then map out the regime in parameter space where this criterion is satisfied for a 16 quantum dot system.

\section{Results}

\begin{figure}
    \centering
    \includegraphics[width=0.9\linewidth]{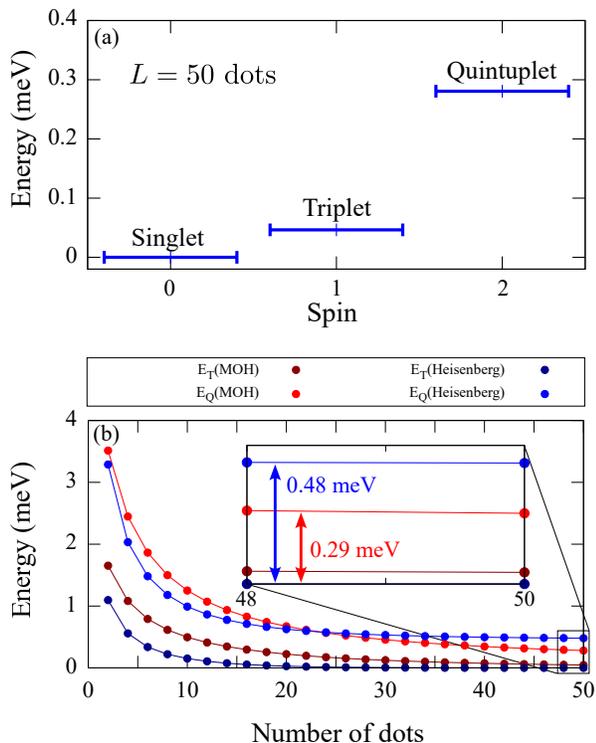}
    \caption{Parameters used are in Table \ref{tab:params}. (a) Low energy spectrum of a chain of 50 quantum dots using the multi-orbital Hubbard model (MOH). (b) Low energy spectrum of a quantum dot array as a function of array size using various models. $E_T$ and $E_Q$ denote triplet and quintuplet energy respectively. All energies are shifted so that the singlet energy, which is not shown, is zero. The inset shows an enlarged section of the plot from $L=48$ to $50$ dots.}
    \label{fig:50_dots}
\end{figure}

One of the ways to determine the spin-1 chain characteristics of the quantum dot array is to observe a quasidegenerate singlet-triplet ground state with a gap that separates the ground state from the quintuplet state in the low energy spectrum. Similar to the multi-orbital Hubbard spectrum for two quantum dots in Fig. \ref{fig:2dot_spec_comparison}(a), the spectrum of a chain of many dots in Fig. \ref{fig:50_dots}(a) also consists of a ground state singlet followed by triplet states and quintuplet states. The difference is that unlike the spectrum of two quantum dots, the singlet and triplet states in the spectrum of the large chain are almost degenerate with a splitting of $0.05$ meV and are separated by a spectral gap from the quintuplet state. The almost degenerate singlet-triplet states along with the spectral gap are indications of the existence of Haldane spin-$\frac{1}{2}$ quasiparticles at the edges. 
To demonstrate this point further, we show the energy of singlet and triplet states as a function of system size in \ref{fig:50_dots}(b).
We see that singlet and triplet become almost degenerate while the singlet-quintuplet energy gap approaches a value of about $0.29$ meV.

\begin{figure}
    \centering
    \includegraphics[width=0.9\linewidth]{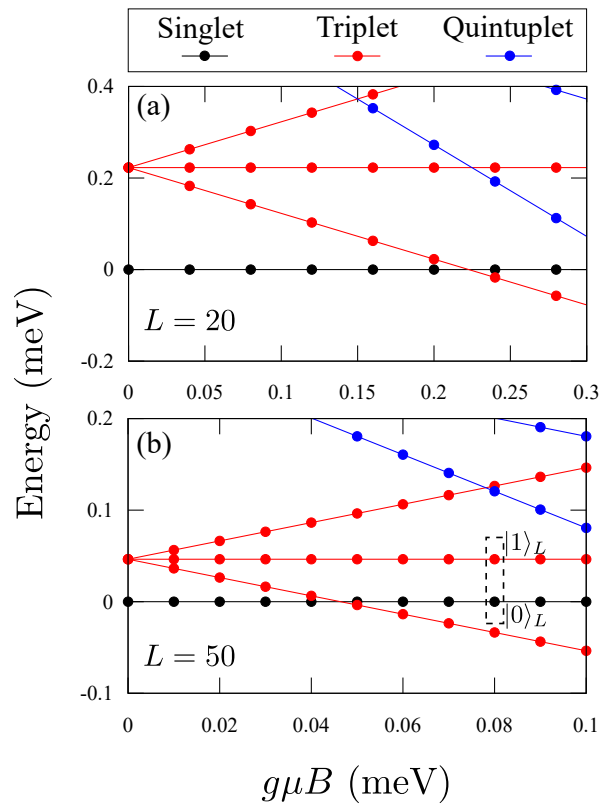}
    \caption{Multi-orbital Hubbard spectra as a function of magnetic field $g\mu B$ for $L=20$ (a) and $L=50$ (b) quantum dot arrays. The logical qubit states are highlighted in the dashed square. }
    \label{fig:b_field}
\end{figure}

The same behaviour is observed in the spectrum of the Heisenberg spin-1 chain in Fig. \ref{fig:50_dots}(b) where the spectral gap is $0.45$ meV. This spectral gap is known as the Haldane gap. Though the spectral gap for the multi-orbital Hubbard model is only about 65\% of the spectral gap in the Heisenberg spin-1 chain spectrum, this level of agreement is to be expected considering the fact that the ground state of a two dot multi-orbital Hubbard model given these parameters is in about $70\%$ agreement with the ground state of a two site Heisenberg spin-1 chain as seen in the overlap integral which was calculated in Ref. \cite{manalo_prb_2021}.
Next, we apply a magnetic field by adding Zeeman energy to demonstrate that the quantum dot array behaves the same way as two Haldane spin-$\frac{1}{2}$ quasiparticles would in a magnetic field and to also show that the $S^z=0$ triplet and singlet can be isolated for use as a qubit basis. The spectra for $20$ and $50$ quantum dot arrays as a function of applied magnetic field are shown in Fig. \ref{fig:b_field}.

In both the $L=20$ and $L=50$ cases, the Zeeman splitting between the triplet components increases as a function of magnetic field while the singlet remains unaffected, which is also the case for two coupled spin-$\frac{1}{2}$ particles. With the inclusion of the Zeeman splitting of the quintuplet components, this system behaves as a Heisenberg spin-1 chain would under a magnetic field.

For qubit operation, it is necessary for the $S^z=\pm 1$ triplet components to split away far enough to isolate the $S^z=0$ triplet and singlet before the $S^z=-2$ quinutplet crosses the $S^z=0$ triplet. This does not happen with the $L=20$ chain as seen in Fig. \ref{fig:b_field} (a) where at about $g\mu B=0.24$ meV, the $S^z=-1$ triplet begins to cross below the singlet, but by then the lowest energy quintuplet already crossed below the $S^z=0$ triplet. For the $L=50$ chain, the singlet-triplet splitting at zero field is small enough such that isolation of the zero singlet and triplet occurs before any quintuplet crossing occurs. At about $g\mu B=0.07$ meV as seen in Fig. \ref{fig:b_field} (b), the nonzero triplets isolate the qubit basis before the lowest energy quintuplet crosses even the $S^z=+1$ triplet. Since it is useful to have a larger spacing between the qubit basis and the nonzero triplets, we can use a magnetic field of $0.08$ meV for qubit operation where the high energy triplet and the low energy quintuplet begin to cross.

It is also useful to construct a synthetic spin-1 chain with other parameters. These multi-orbital Hubbard parameters depend on the material, quantum dot As concentration, interdot distance and quantum dot size. Varying these parameters would vary the spectral gap since $J_1 \propto \frac{t^2}{U+\frac{J_{1/2}}{2}-V}$ \cite{blazej_thesis_2018}, hence it is important to find which parameters would yield a synthetic spin-1 chain.

\begin{figure}
    \centering
    \includegraphics[width=\linewidth]{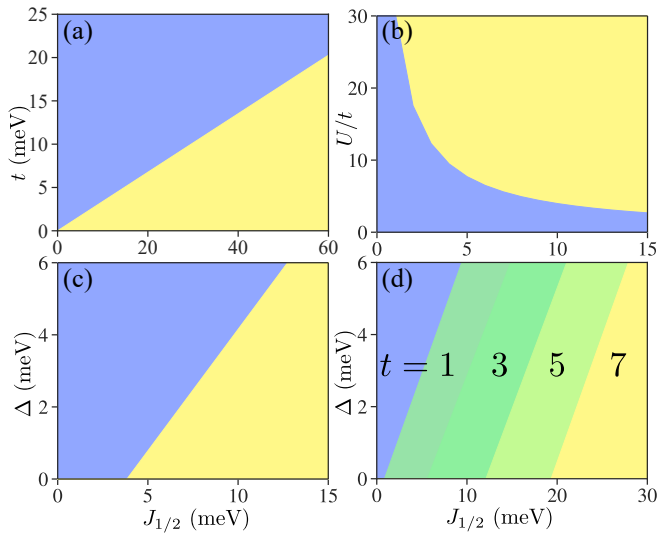}
    \caption{Spin-1 chain spectrum criterion as a function of various multi-orbital Hubbard parameters for an array of two quantum dots. Yellow (or green) region is where the criterion
    is satisfied and blue is otherwise. (a) is a diagram of $t$ and $J_{1/2}$, (b) is a diagram of $U/t$ and $J_{1/2}$, (c) shows $\Delta$ versus $J_{1/2}$ and (d) shows $\Delta$ versus $J_{1/2}$ at different values of $t$, where the all values of $t$ are in units of meV.
    }
    \label{fig:2_dots}
\end{figure}

\begin{figure}
    \centering
    \includegraphics[width=\linewidth]{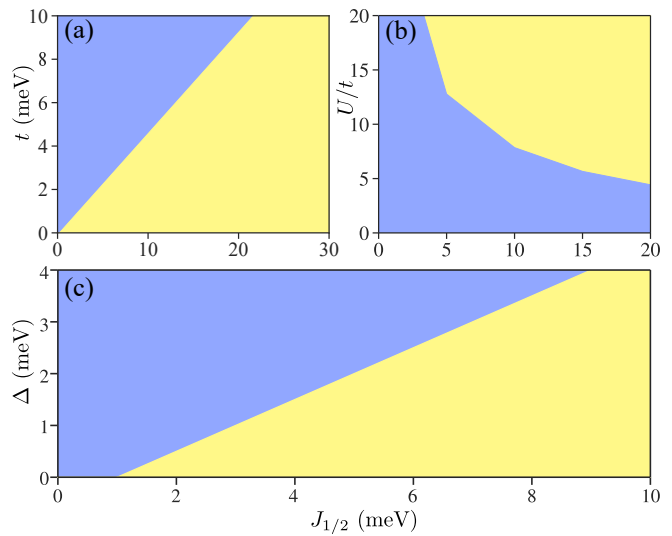}
    \caption{Spin-1 spectrum criterion as a function of various multi-orbital Hubbard parameters for an array of 16 quantum dots. Yellow region is where the criterion
    is satisfied and blue is otherwise. All parameters are the same as the ones used in Figs. \ref{fig:50_dots} and \ref{fig:b_field} except for those that are varied.}
    \label{fig:16_dots}
\end{figure}

We map out regions in parameter space for 2 dot and 16 dot arrays where the Hamiltonian (\ref{eq:hk_multidot}) produces a spin-1 chain spectrum, that is, regions where the spectrum consists of a singlet ground state, a triplet first excited state and a quintuplet second excited state. In both Figs. \ref{fig:2_dots} and \ref{fig:16_dots}, there are clear regions in parameter space where a spin-1 spectrum is produced as opposed to random points sporadically dispersed. We decided to omit the parameter $V$ from the diagrams because the term containing this factor only contributes a constant shift to the low energy spectrum due to all of the orbitals in these states having single occupation. 
In Fig. \ref{fig:2_dots}(a) we see a $t$ varying linearly with $J_{1/2}$ at the boundary. Moreover, Fig. \ref{fig:2_dots}(b) shows a $\frac{1}{J_{1/2}}$ dependence of $U$ at the boundary, which is expected because of the linear dependence of $t$ on $J_{1/2}$ at the boundary in Fig. \ref{fig:2_dots}(a). In Fig. \ref{fig:2_dots}(c), we see a linear dependence between $\Delta$ and $J$ at the boundary. Furthermore, varying $t$ at different cross sections of the $\Delta$-$J_{1/2}$ plane as shown in Fig. \ref{fig:2_dots}(d), does not vary the slope of the boundary. However, the $J_{1/2}$-intercept increases with $t$.

Since the singlet-triplet qubit requires large chains, we also map out regions in parameter space where a spin-1 chain spectrum is produced for a 16 dot system in Fig. \ref{fig:16_dots}. Similar trends to the 2 dot diagrams are seen in the 16 dot diagrams. For instance, linear dependence of $t$ on $J_{1/2}$ at the boundary where the spin-1 chain criterion is satisfied is shown in Fig. \ref{fig:16_dots}(a). The $\frac{1}{J_{1/2}}$ dependence in Fig. \ref{fig:16_dots}(b) is also seen as well as the linear dependence of $\Delta$ on $J_{1/2}$ in Fig. \ref{fig:16_dots}(c).

The two lowest energy eigenstates for a single quantum dot with two electrons in the $p$-shell are a triplet ground state and a singlet first excited state separated by an energy $E_{s1}=\frac{3J_{1/2}-\sqrt{J_{1/2}^{2}+(4\Delta)^{2}}}{4}$ \cite{manalo_prb_2021}. If another quantum dot is placed beside the first one, the orbitals in different dots are coupled by the hopping term $t c_{i\alpha}^\dag c_{i+1\alpha}$. We see the effect of hopping in the spectrum of $4$ electrons on two quantum dots in the splitting of the singlet and triplet double quantum dot states, where in this case, the singlet is the ground state and the first excited state is a triplet. In the regime where the hopping term in (\ref{eq:hk_multidot}) is weak, the singlet triplet splitting is proportional to the effective Heisenberg spin-1 coupling $J_1 = 2t^2/\left( U + J_{1/2}/2-V \right)$. This introduces the condition  $J_1 < E_{s1}$ which can be interpreted as the values of $t$ which conserve the spin-1 character of each quantum dot in the array. This condition can shed light on Figs. \ref{fig:2_dots} and \ref{fig:16_dots}. 

Increasing $J_{1/2}$ increases the singlet-triplet splitting for a single quantum dot, protecting the spin-1 character of each dot against perturbations according to the analytic expression of $E_{s1}$. This behavior is observed in Figs. \ref{fig:2_dots} and \ref{fig:16_dots}, where the Haldane phase is favored whenever $J_{1/2}$ is increased. On the other hand, an increase in the hopping energy $t$ mixes the single quantum dot ground and excited states, destabilizing Haldane phase , which is also observed in Figs. \ref{fig:2_dots}(a) and \ref{fig:16_dots}(a). Similarly, increasing the $p$-shell splitting $\Delta$ decreases the singlet-triplet gap in a single quantum dot, eventually producing  a ground state that is a singlet instead of a triplet in the case when $E_{s1} < 0$, losing the spin-1 behavior of each quantum dot. 

In Figs. \ref{fig:2_dots}(c) and \ref{fig:16_dots}(d) the competition between $J_{1/2}$ and $\Delta$ terms can be seen directly. Particularly, in Fig. \ref{fig:2_dots}(d) the combined effect of varying $\Delta$ with the hopping term $t$ and $J_{1/2}$ is shown. When the hopping energy $t$ is increased, $J_{1/2}$ must also increase even for negligible $\Delta$ in order to stabilize the Haldane phase. This is the reason behind the increasing of the $J_{1/2}$-intercept as $t$ increases.

\section{Conclusion}
We presented here the steps enabling the microscopic design of a topologically protected singlet-triplet qubit in an InAsP quantum dot array embedded in an InP nanowire.
 The multi-orbital Hubbard model derived from microscopic calculations is used to describe the singlet-triplet qubit. A degenerate singlet-triplet ground state followed by a spectral gap separating the ground state from the quintuplet state is observed in the low energy spectrum of the multi-orbital Hubbard chain. This same behaviour which is also observed in the spectrum of a Heisenberg spin-1 chain indicates the existence of spin-$\frac{1}{2}$ quasiparticles at the edges of the chain. Further indication of the existence of spin-$\frac{1}{2}$ quasiparticles is the behaviour of the low energy spectrum as a function of applied magnetic field. Despite the system being a chain of synthetic spin-1s constructed using an InAsP quantum dot array, the magnetic field dependence of the spectrum is the same as that of two spin-$\frac{1}{2}$ quasiparticles. The external magnetic field also allows the $S^z=0$ triplet and singlet states to serve as the qubit basis by isolating those states from the other $S^z=\pm 1$ components of the triplet and the quintuplet states. For the design of the qubit, the regions in parameter space where the low energy spectrum of the Heisenberg spin-1 chain is reproduced with the multi-orbital Hubbard model are mapped. The finite spectral gap gives the qubit potential to be robust against perturbations.

\begin{acknowledgments}
JM and DM thank Y. N\'u\~nez-Fern\'andez for useful discussions. This
research was supported by NSERC QC2DM Strategic Grant
No. STPG-521420, NSERC Discovery Grant No. RGPIN2019-05714, and University of Ottawa Research Chair in
Quantum Theory of Quantum Materials, Nanostructures, and
Devices and Compute Canada 
for computing resources.
\end{acknowledgments}
\providecommand{\noopsort}[1]{}\providecommand{\singleletter}[1]{#1}%
%


\end{document}